\documentstyle[12pt,aasms4]{article}
 
\slugcomment{December 17, 1998; to appear in Astrophysical Journal
(letters)} 
 
\begin{document}
 
\title{The Radio-to-Submm Spectral Index as a Redshift Indicator}
 
\author{C. L. Carilli and Min Su Yun}
\affil{National Radio Astronomy Observatory, P.O. Box O, Socorro, NM, 87801 \\}
\authoremail{ccarilli@nrao.edu, myun@nrao.edu}
 
\begin{abstract}

We present models of the 1.4 GHz to
350 GHz  spectral index, $\alpha^{350}_{1.4}$,
for starburst galaxies as a function of redshift. The
models  include a semi-analytic formulation, based on the well
quantified radio-to-far infrared correlation for low redshift star
forming galaxies, and an empirical formulation, based on the 
observed spectrum of the starburst galaxies M82 and Arp 220. 
 We compare the models to the observed 
values of $\alpha_{1.4}^{350}$ for starburst galaxies at low and high
redshift. We find reasonable agreement between the models and the
observations, and in particular, that an observed spectral index of 
$\alpha_{1.4}^{350}$ $\ge$ +0.5 indicates that the target
source is likely to be at high redshift, z $\ge$ 1. 
The evolution of $\alpha^{350}_{1.4}$
with redshift is mainly due to the very steep rise in
the Raleigh-Jeans portion of the thermal dust spectrum shifting
into the 350 GHz band with increasing redshift.  
We also discuss situations where this relationship could be violated.  
We then apply our models to examine the putative identifications of
submm sources in the Hubble Deep Field, and conclude that 
the submm sources reported by Hughes et al. are likely to be at
high redshifts, $\rm z \ge 1.5$.  
\end{abstract}
 

\section {Introduction}

Detecting submm continuum emission from objects at z $\ge$ 2
has revolutionized our understanding of galaxies at high redshift
(Hughes et al. 1998, Ivison et al. 1998, Smail, Ivison, and Blain
1997, Eales et al. 1998, Barger et al. 1998). 
The emission is thought to be thermal emission from warm dust, with
implied dust masses $\ge$ 10$^8$ M$_\odot$. A number of these submm
sources  have also been detected in CO emission with implied molecular
gas masses $\ge$ 10$^{10}$ M$_\odot$ (Brown \& Vanden Bout 1991,
Barvainis et al. 1994, Ohta et al. 1996, Omont et al. 1996a,
Guilloteau et al. 1997, Frayer et al. 1998). The large reservoirs of
warm gas and dust in these systems has led to the hypothesis that
these are starburst galaxies, with massive star formation rates $\ge$
100 M$_\odot$ year$^{-1}$ (Hughes \& Dunlop 1998). In some cases,
there may be an associated active galactic nucleus (AGN), leading to
questions about the dominant dust heating mechanism -- star formation,
or AGN, or both (cf. Sanders \& Mirabel 1996, Downes \& Solomon 1998,
Smith  et al. 1998)?

A well studied phenomenon in nearby star forming galaxies is
the radio-to-far IR correlation, ie. the
tight correlation found between the radio continuum emission and
the thermal dust emission  (Condon 1992, Helou and Bicay 1993). 
The radio continuum emission is thought to be synchrotron radiation
from relativistic electrons spiraling in interstellar magnetic
fields. The standard explanation for 
the radio-to-far IR correlation involves relativistic
electrons accelerated in supernova remnant shocks, and  dust
heated by the interstellar radiation field. 
Both quantities are then functions of the massive star formation rate
(Condon 1992, Cram et al. 1998, Yun et al. 1998, 
Gruppioni, Mignal, and Zamorani 1998), although the
the detailed physical processes giving rise to the tight correlation
remain enigmatic. If the radio-to-far IR correlation is
independent of redshift, then the sharp rise in the
the Raleigh-Jeans portion of the thermal dust spectrum shifting
into the 350 GHz band with increasing redshift
implies that the observed spectral index between radio and submm
frequencies should evolve strongly with redshift (Hughes et
al. in preparation).

In this paper we explore the possibility of using the radio-to-submm
spectral index as a redshift indicator for star forming galaxies.
Models of the expected spectral index between 1.4 GHz and 350 GHz (850
$\mu$m), $\alpha_{1.4}^{350}$, are presented based on the standard
relationships derived for nearby star forming galaxies, and on the
observed spectra of two `canonical' starburst galaxies, M82 and Arp
220. We present a simple analytic expression relating redshift to 
$\alpha_{1.4}^{350}$, and we compare the models to the observed 
values of $\alpha_{1.4}^{350}$ for starburst galaxies at low and high
redshift. We find reasonable agreement between the models and the
observations, and in particular, that an observed spectral index of 
$\alpha_{1.4}^{350}$ $\ge$ +0.5 indicates that the target
source is likely to be at high redshift,  z $\ge$ 1. 
We discuss possible `confusing' mechanisms which could
complicate the analysis, such as radio emission driven by an
AGN, dust heated solely by a radio quiet AGN, 
and free-free absorption at low radio frequencies. 
We then apply our models to examine the putative identifications of
submm sources in the Hubble Deep Field (HDF).  

\section{Analysis}

Condon (1992) presents  semi-analytic, linear relationships
between  the massive star formation rate and the radio and far IR
emission from active star forming galaxies. We have used these
relationships to  derive a simple relationship between redshift and 
$\alpha_{1.4}^{350}$, by making a
simplifying assumption that the source spectrum can be characterized by
two power-law spectra, one at low frequencies (observing frequency
$\le$ 30 GHz) with a spectral index $\alpha_{\rm radio}$, and one at
high frequencies (observing frequency between 230 GHz and
850 GHz), with index $\alpha_{\rm submm}$. We define spectral index in
terms of frequency, $\nu$, and the observed flux density, S$_\nu$, as:
S$_\nu$ $\propto$ $\nu^{\alpha}$. 
Equation~21 in Condon (1992) relating radio synchrotron luminosity
to massive  star formation rate can be written in terms of 
observed flux density as:
$$ \rm S_{radio} = 4\times10^{28} [{{(1 + z)^{1+\alpha_{radio}}}\over{4
\pi D_L^2}}] [{{\nu_{radio}}\over{1.4~GHz}}]^{\alpha_{radio}} \times
SFR ~~~ erg~ cm^{-2}~ s^{-1}~ Hz^{-1} ~~~~~~(1)$$
where S$_{\rm radio}$ is the observed radio flux density
due to synchrotron emission,
$\nu_{\rm radio}$ is the observing frequency,
D$_{\rm L}$ is the source
luminosity distance, and SFR is the star
formation rate for stars with masses $\ge$ 5 M$_\odot$, in units of
$M_\odot$ year$^{-1}$. The equation relating submm emission from warm
dust to massive  star formation rate can be derived using the 
Condon's Equation~26, assuming a spectrum of the form observed
for nearby active star forming galaxies such as M82:
$$ \rm S_{submm} = 1\times10^{28} [{{(1 + z)^{1+\alpha_{submm}}}\over{4
\pi D_L^2}}] [{{\nu_{submm}}\over{350~ GHz}}]^{\alpha_{submm}} \times
SFR ~~~ erg~ cm^{-2}~ s^{-1}~ Hz^{-1} ~~~~(2)$$
where S$_{\rm submm}$ is the submm flux density due to thermal dust
emission, and $\nu_{\rm submm}$ is the observing frequency.
We use H$_o$ = 75 km s$^{-1}$ Mpc$^{-1}$ and q$_o$ = 0.5 where
required. 

Taking the ratio of these two expressions, it is straight forward to
show that the spectral index between 1.4 GHz and 350 GHz,
$\alpha_{1.4}^{350}$, behaves as a function of redshift, z, as:
$$\rm  \alpha_{1.4}^{350} = -0.24~ - ~ [0.42 \times (\alpha_{radio} -
\alpha_{\rm submm}) \times log(1 + z)] ~~~~(3)$$
This relationship is plotted in Figure 1.
For $\alpha_{\rm radio}$ we adopt the standard value in Condon (1992) of
$-$0.8. For $\alpha_{\rm submm}$ we use values of +3.0 (solid
curve) and +3.5 (short dashed curve). Note
that the observed spectral indices  between 270 GHz and 850 GHz
for M82 and Arp 220 are +3.4 and +3.0, respectively.

We have also derived values of $\alpha_{1.4}^{350}$ as a function of
redshift for two `canonical' starburst 
galaxies, M82 and Arp 220. These two galaxies have 
well sampled spectra over the entire frequency range from the radio
into the optical (Klein et al. 1988, Scoville et al. 1991). 
The $\alpha_{1.4}^{350}$ models were derived by fitting 
accurate polynomials to the observed data from 1.4 GHz to 22 GHz, and
from 230 GHz to 20000 GHz. These results are shown in Figure~1 as a
dotted curve   for M82 and a long dashed curve for Arp 220.

The empirical models for M82 and Arp 220 differ from the simplified two
power-law models in two important ways. First, the observed
spectral indices at zero redshift are typically higher for the
empirical models relative to the two power-law model. This difference
is most likely due 
to a low frequency flattening at 1.4 GHz due to free-free absorption
in the denser HII regions in the galaxy (Condon 1992). For instance,
the observed spectral index for M82 between 1.4 and 5 GHz is $-0.58$,
while the spectral index between 5 GHz and 10.7 GHz is $-0.72$ 
(Klein et al. 1988). 
And second, the two power-law model diverges at
large redshift while the empirical models flatten and eventually
turn-over at z $>$ 7. This effect is due to the fact that the 
thermal spectra peak around 3000 GHz (100 $\mu$m) for a 
dust temperature of 30 K.
An observed frequency of 350 GHz corresponds
to a rest frame frequency of 3000 GHz at z = 7.5, hence at higher
redshift the spectrum has gone `over-the-top' of the thermal peak.

Plotted on Figure 1 are values of  $\alpha_{1.4}^{350}$
for galaxies detected at 350 GHz and 1.4 GHz. The submm data for the
low redshift galaxies are from a survey of nearby active star forming
galaxies using the JCMT (Hughes et al. 1990, Rigopoulou et al. 1996) 
while the submm data for the z $>$ 1 sources are from 
Rowan-Robinson et al. (1993), Isaak et al. (1994), 
Barvainis et al. (1995), Omont et al. (1996b),
Dey et al. (1998), Cimatti et al. (1998),
Ivison et al. (1998ab), Lewis et al. (1998), Eales et al. (1998),
and Kawabe et al. (1998).  All the radio data are from the Very
Large Array (VLA) (see also Fomalont et al. 1991). Overall, the models
appear to define (within the 
errors) the range in observed values of  $\alpha_{1.4}^{350}$ as a
function of redshift. In particular, the galaxies at z $>$ 1.5 have
$\alpha_{1.4}^{350}$ values $\ge$ +0.5, while the low redshift galaxies
have   $\alpha_{1.4}^{350}$ values $\le$ +0.2. 

The scatter in
the data for low redshift galaxies is, again,  due in part 
to variations in free-free absorption 
at 1.4 GHz between sources. In the more extreme cases, the
implied (mean) free-free optical depths are $\ge$ 1 at 1.4 GHz,
implying emission measures $\ge$ 6$\times$10$^6$ $\times$ ($\rm
{T_K}\over{10^4}$)$^{3\over2}$ for the starburst regions 
(Taylor et al. 1998). Note that this phenomenon is relevant only for
low z galaxies since the low frequency turnover rapidly shifts out of the
1.4 GHz band with increasing redshift. 

The scatter in the data at high redshift may
be due, in part, to contamination of the radio emission by an active
nucleus.  The frequency of radio AGNs among an IR selected galaxy
sample is about 10\% for galaxies with $L_{FIR}\ge 10^{11}~L_\odot$
(Yun et al. 1998).
Radio AGN emission will cause the observed value of 
$\alpha_{1.4}^{350}$ to fall below the values predicted for star
forming galaxies.  One clear example of this in Figure~1a is 
the Clover Leaf Quasar, H1413+117 at
z = 2.56 (Barvainis et al. 1995), which has a rest frame radio continuum
luminosity of 1.3$\times$10$^{32}$ ergs s$^{-1}$ Hz$^{-1}$ at 
1.4 GHz, allowing for a  magnification factor of 7.6 
by gravitational lensing. 
Even corrected for lensing, the implied star formation rate 
is unreasonably high (3300 M$_\odot$ year$^{-1}$ using Eq.~21 of
Condon 1992), and the expected 350 GHz flux density
is 230 mJy. The observed 350 GHz flux density is a factor of five lower. 
It is likely that H1413+117 is a 
Fanaroff-Riley Class I (`FRI' = low luminosity) radio galaxy, with a
radio luminosity two times larger than M87. 
One method for separating AGN-driven radio emission from
starburst driven radio emission is sub-arcsecond imaging in the radio
and submm, to look for spatial coincidence of the radio and submm
emission. Note that we have not included the submm detections of 
high redshift Fanaroff-Riley Class II (high luminosity) radio galaxies
in this study, such as  4C 41.17 at z = 3.8 and 1435+635  
at z = 4.25 (Hughes \& Dunlop 1998). The extreme radio powers of these
objects (1000 $\times$ M87) imply $\alpha_{1.4}^{350}$ $\le$ $-$0.5,
which is  off the bottom of Figure 1, and they can be unambiguously
recognized as such.

A third possible uncertainty in Figure 1 can arise due to
gravitational lensing. A 
number of the high redshift sources are known to be gravitational
lenses (cf.  Barvainis 1998, Blain 1998). 
If the radio continuum and submm emission are 
roughly co-spatial, as would be the case for a starburst galaxy, then  
gravitational lensing will not affect the $\alpha_{1.4}^{350}$ values.
However, if the radio emission is distributed differently than the
submm emission, as could occur for a radio loud AGN, then
differential magnification by a lens could lead to significant
variations in the observed $\alpha_{1.4}^{350}$. 

The general agreement between the models and the data in Figure 1
arises mainly from the very sharply rising submm
spectrum of thermal dust emission ($\alpha_{\rm submm}$ $\ge$ +3). This
sharp rise in the submm spectrum, coupled with the 
large `lever-arm' in frequency between 1.4 GHz and 350 GHz, 
can mitigate  the uncertainties in the radio 
spectrum, such as free-free absorption, or even low
luminosity radio AGN emission (cf. Schmitt et al. 1997). 
A submm frequency of 350 GHz is 
a good choice for this study, since it is close to the minimum
on the Raleigh-Jeans part of the thermal dust spectrum for low z
galaxies (cf. Condon 1992), for which the  
value of $\alpha_{1.4}^{350}$ should be close to zero,
and it does not reach the peak in the dust emission spectrum
until z $\approx$ 7 (see above), where $\alpha_{1.4}^{350}$ $\ge$~1.
Since the strength of the method lies in the steep rise in the
dust spectrum, van der Werf et al. (1999) have recently performed an
analogous  analysis using the optical-to-far IR spectral index as a
redshift indicator. One problem that occurs in the optical is
confusion. A typical submm error box of 3$''$ has
a 50$\%$ chance of containing a `random'
optical galaxy at the limit of the HDF (I$_{814}$ $\le$ 29;
Williams et al. 1996), while at 1.4 GHz
the probability is only 1$\%$ for finding a $\ge$ 10 $\mu$Jy
source in the error box (Langston et al. 1990). 

\section{Discussion}

Figure 1 can be considered in two ways.
The first is as a redshift indicator for star forming galaxies.
In this regard,
the most important conclusion that can be reached from Figure 1 is
that a value of $\alpha_{1.4}^{350}$ $\ge$ +0.5 indicates 
that the source is likely to be at high redshift, z $\ge$ 1.

One possible mechanism that could lead  to larger
$\alpha_{1.4}^{350}$ values for 
starburst galaxies than predicted by 
the models based on the radio-to-far IR correlation would be to
`quench' the radio continuum emission associated with star
formation through inverse Compton losses off the microwave
background radiation field. However, this mechanism is likely to
become  important only at very high redshift (z $\ge$ 6).
The energy density in the microwave background increases
as (1+z)$^4$, which, at z = 6, corresponds to the energy density 
in a magnetic field of about
100 $\mu$G -- comparable to the expected interstellar
magnetic fields in starburst nuclei (Condon et al. 1991,
Carilli et al. 1998). The lifetime for a relativistic particle
radiating at a rest frequency of 10 GHz  (= 1.4 GHz observed
frequency), is then 0.5 Myr. 

Contamination of the radio continuum emission by a low luminosity 
radio AGN could lead to an ambiguity for 
values  of  $\alpha_{1.4}^{350}$ $\le$ +0.2, such that the source is
either a starburst galaxy at low redshift, 
or a radio loud AGN at higher redshift.
The amount of contamination by radio loud AGN in a given galaxy sample
will  depend on the relative
cosmic density of active star forming galaxies versus radio loud AGN,
and on the flux density limits of the observations. Assuming
sensitive observations can be made of galaxies with star formation
rates of order 10 M$_\odot$ year$^{-1}$ out to high redshift (see
below), then the expected contamination by FRI-class radio galaxies
should be $\le 30\%$, based on galaxy populations at low
redshift (Yun et al. 1998, Osterbrock 1989, 
Hammer et al. 1995, Richards et al. 1998).  Of course, this fraction
could change with redshift (Gruppioni et al. 1998).  

The second use for Figure 1 is as an `AGN indicator'. Given a
value of  $\alpha_{1.4}^{350}$, and an independent estimate of the 
source spectral redshift, if the source lies well below the curves in
Figure 1, then it is likely the source has a radio loud AGN component. 
Again, H1413+117 is a good example of this. 
Theoretically, a source can appear 
well above the curves in Figure 1 if the dust heating
mechanism is entirely due to a radio quiet AGN.
Thus far no examples of this latter type
have been found, but  current limits on a few sources cannot 
preclude such a situation. 

We can apply the analysis in Figure 1 to address the identification
of the submm sources detected in the Hubble Deep Field (HDF)
by Hughes et al. (1998). 
Because there are of order 10 optical galaxies within a given 
SCUBA beam at 350 GHz (FWHM = 15$''$), unique identification
of the submm sources is difficult from the astrometry
alone.  Hughes et al. have argued that the submm sources 
are at $\rm z\ge1$ based on the submm spectral shape.
Using the deep radio imaging data from the VLA, 
Richards (1998) has argued that there is a $6''$ offset between the
submm frame with respect to the radio and optical frame, and that the
sources HDF850.3 \& HDF850.4 could be identified with bright optical
galaxies at $\rm z\sim 0.5$ instead.  As shown in Figure~1b,
however, the derived radio-to-submm spectral index for
the three proposed identifications by Richards (shown as filled
circles) are much too large to be consistent with their redshifts.  
Reversing the argument, the
maximum expected 350 GHz fluxes for the two bright 
$\rm z\sim 0.5$ radio sources are 0.21 and 0.44 mJy, factors of 
14 and 5.2 too small compared to the  observed values.

Conversely, if the SCUBA astrometric accuracy is better than 6$''$, 
only upper limits exist for the 1.4 GHz radio flux ($3\sigma
=23~\mu$Jy).  The resulting {\it lower} limits to 
$\alpha_{1.4}^{350}$ for the five HDF submm sources
are shown in Figure~1b using the redshifts estimated by Hughes
et al.  The derived $\alpha_{1.4}^{350}$ is larger than +0.8 
in all five cases, and therefore these sources are likely to be
located at $\rm z\ge 1.5$, more in-line
with the approximate redshifts estimated for the sources based solely
on the submm spectral indices by Hughes et al. (1998). 
Further, the brightest submm source HDF850.1 has 
$\alpha_{1.4}^{350}>$ +1.0, larger than any previously detected
submm source (see Figure~1a), suggesting $\rm z \ge 3$.
For comparison, the z=4.7 QSO BR1202-0725
has $\alpha_{1.4}^{350}=$ +0.92.

Recent infrared, submm, and radio observations suggest that 
optical studies of galaxies at high redshift may miss a 
significant population of massive star forming galaxies due to dust
obscuration, and that the cosmic  star
formation rate may be a factor three or more larger at z $\ge$ 1 than
estimated from optical studies (Flores et al. 1998, 
Hughes et al. 1998). Sensitive submm 
observations using bolometer arrays, and sensitive radio observations
using the VLA, are typically 
limited to sources with flux densities $\ge$ 1 mJy at
350 GHz, and $\ge$ 40 $\mu$Jy at 1.4 GHz (Richards et al. 1998),
corresponding to sources with star formation rates $\ge$ 100 M$_\odot$
year$^{-1}$ at z $\approx$ 2. The large millimeter array currently being
designed for the high desert in Chile will have a 350 GHz sensitivity
of about 10 $\mu$Jy in one hour.
This will be easily adequate to detect  galaxies with star
formation rates of  order 10 M$_\odot$ year$^{-1}$ 
at z = 3 (Brown 1996), for which the expected flux density at 350 GHz
is 100 $\mu$Jy. Likewise, the up-graded VLA
will have  sub-$\mu$Jy sensitivity at 1.4 GHz and 2.3 GHz for 
full synthesis observations.
This should be adequate to image these same
galaxies, for which the expected 
flux density at 1.4 GHz is a few $\mu$Jy, 
with sub-arcsecond resolution. Hence, in the near future we
should be able to extend radio and submm studies of star
formation to `normal' galaxies at substantial look-back times. 

\vskip 0.2truein 

We would like to thank F. Owen, D. Hughes, E. Richards, and K. Menten
for useful discussions, and R. Ivison  for communicating his
unpublished submm and radio results. This research made use of the
NASA/IPAC Extragalactic Data Base (NED) which is operated by the Jet
propulsion Lab, Caltech, under 
contract with NASA. The National Radio Astronomy Observatory (NRAO) is
a facility of the National Science Foundation, operated under
cooperative agreement by Associated Universities, Inc.

\vfill\eject

\centerline{\bf References}

Barger, A.J., Cowie, L.L., Sanders, D.B., Fulton, E., Taniguchi, Y.,
Sato, Y., Kawabe, K., \& Okuda, H. 1998, Nature, 394, 248

Barvainis,R., Tacconi,L., Antonucci,R., Alloin,D., \&
Coleman,P. 1994, Nature, 371, 586

Barvainis,R., Antonucci,R., Hurt,T., Coleman,P., \& Reuter,H.-P.
1995, ApJ, 451, L9

Barvainis, R. 1998, in {\sl Highly Redshifted Radio
Lines}, eds. Carilli, Radford, Menten, \& Langston, (San Francisco:
PASP), p. 40

Blain, A.W. 1998, MNRAS, 297, 511

Brown, R.L. 1996, in {\sl Cold Gas at High Redshift}, eds. Bremer, van
der Werf, R\"ottgering, \& Carilli (Dordrecht: Kluwer), p.  411

Brown, R.L., \& Vanden Bout, P.A. 1991, AJ, 102, 1956

Carilli, C.L., Wrobel, J.M., \& Ulvestad, J.S. 1998, AJ, 115, 928

Cimatti, A., Andreani, P., Rottgering, H., \& Tilanus, R. 1998,
Nature, 392, 895

Condon, J.J. 1992, ARAA, 30. 575

Condon, J.J., Huang, Z.-P., Yin, Q.F., \& Thuan, T.X. 1991, ApJ  378,
65 


Cram, L., Hopkins, A., Mobasher, B., \& Rowan-Robinson, M. 1998,
Ap.J., 508, 155

Dey, A., Graham, J.R., Ivison, R.J., Smail, I., \& Wright, G.S. 1998,
ApJ, submitted

Downes, D., \& Solomon, P.M. 1998, ApJ., in press,

Eales, S., Lilly, S., Gear, W., Dunne, L., Bond, J.R., Hammer, F.,
Le Fevre, O., \& Crampton, D. 1998, ApJ, submitted

Flores, H., Hammer, F., Thuan, T., Cesarsky, C., Desert, F., Omont,
A., Lilly, S., Eales, S., Crampton, D., \& Le Fevre, O. 1998, Ap.J.,
in press 

Fomalont, E.B., Windhorst, R.A., Kritian, J.A., and Kellermann,
K.I. 1991, A.J., 102, 1258

Frayer, D.T., Ivison, R.J., Scoville, N.Z., Yun, M., Evans, A.S.,
Smail, Ian, Blain, A.W., \& Kneib, J.-P. 1998, Ap.J. (letters), 506, 7 

Hammer, F., Crampton, D., Lilly, S., Le Fevre, O., and Kenet, T. 1995,
MNRAS, 276, 1085

Gruppioni, C., Mignoli, M., \& Zamorani, G. 1998, MNRAS, in press

Guilloteau,S., Omont,A., McMahon,R.G., Cox,P., \& Petitjean,P.
1997, A\&A, 328, L1


Helou, G. and Bicay, M.D. 1993, Ap.J., 415, 93

Hughes, D.H., Gear, W.K., \& Robson, E.I. 1990, MNRAS, 244, 759

Hughes, D.H. \& Dunlop, J.S. 1998, in {\sl Highly Redshifted Radio
Lines}, eds. Carilli, Radford, Menten, \& Langston, (San Francisco:
PASP), p. 103

Hughes, D.H., et al. 1998, Nature, 394, 241

Isaak, K.G., McMahon, R.G., Hills, R.E., \& Withington, S. 1994,
MNRAS, 269, L28

Ivison, R.J.,  Smail, Ian, Le Borgne, J.-F.,  Blain, A.W., Kneib,
J.-P., Bezecourt, J., Kerr, T.H., \& Davies, J.K. 1998a,  MNRAS, 298, 583

Ivison, R.J. et al. 1998b, in preparation

Kawabe, R., Kohno, K., Ohta, K., \& Carilli, C. 1998, in {\sl Highly
Redshifted Radio Lines}, eds. Carilli, Radford, Menten, \& Langston,
(San Francisco: PASP), p. 48

Klein, U., Wielebinski, R., \& Morsi, H.W. 1988, A\&A, 190, 41

Langston, G.I., Conner, S.R., Heflin, M.B., Lehar, J., \& 
Burke, B.F.,  1990, Ap.J., 353, 34

Lewis, G.F., Chapman, S.C., Ibata, R.A., Irwin, M.J., \& Totten,
E.J. 1998, ApJ, 505, L1


Ohta, K., Yamada, T., Nakanishi, K., Kohno, K., Akiyama, M., \&
Kawabe, R. 1996, Nature, 382, 426

Omont, A., Petitjean, P., Guilloteau, S., McMahon, R.G., Solomon,
P.M., \& Pecontal, E. 1996a, Nature, 382, 428

Omont, A., McMahon, R.G., Cox, P., Kreysa, E., Bergeron, J., Pajot, F.,
\& Storrie-Lombardi, L.J. 1996b, A\&A, 315, 1


Osterbrock, D.E. 1989, {\sl Astrophysics of Gaseous Nebulae and Active
Galactic Nuclei}, California: University Science Books

Richards, E.A., Kellermann, K.I., Fomalont, E.B., Windhorst, R.A., \&
Partridge, R.B. 1998, AJ, 116, 1039

Richards, E.A. 1998, Ap.J. (letters), submitted

Rigopoulou, D., Lawrence, A., \& Rowan-Robinson, M. 1996, MNRAS, 278, 1049 

Rowan-Robinson, M. et al. 1993, MNRAS, 261, 513 

Sanders, D.B. \& Mirabel, I.F. 1996, ARAA, 34, 749

Schmitt, H., Kinney, A., Calzetti, D., \& Storchi-Bergmann, T. 1997,
AJ, 114, 592

Scoville, N.Z., Sargent, A.I., Sanders, D.B., \& Soifer, B.T.
1991, ApJ, 366, L5

Smail, I., Ivison, R.J., \& Blain, A.W., 1997, Ap.J., 490, L5

Smith, Hardin E., Lonsdale, Carol J., Lonsdale, Colin J., and Diamond,
P.J. 1997, Ap.J. (letters), 493, 17

Taylor, G.B., Silver, C., Ulvestad, J., \& Carilli, C.L. 1998, 
Ap.J., submitted

van der Werf, P.P., Clements, D.L., Shaver, P.A., \& Hawkins,
M.R. 1999, A\&A, in press

Williams, R. et al. 1996, AJ, 112, 1335

Yun, M.S., Reddy, N., \& Condon, J.J. 1998, in preparation.

\vfill\eject

\centerline{Figure Caption}

{\bf Figure 1:} (a) Two power-law models (Eq.~3) for the 1.4 GHz 
to 350 GHz spectral index, $\alpha_{1.4}^{350}$, for star 
forming galaxies as a function of redshift are shown with a
solid curve ($\alpha_{\rm submm}=$ +3.0) and a short dashed curve
($\alpha_{\rm submm}=$ +3.5).  The assumed radio spectral index is
$-$0.8.  The dotted and long-dashed curves show the expected values of
$\alpha_{1.4}^{350}$ versus redshift based on the observed spectra of
the star forming galaxies M82 and Arp 220, respectively. 
Also plotted in filled squares are the submm data for the
low redshift galaxies from a survey of nearby active star forming
galaxies using the JCMT (Hughes et al. 1990, Rigopoulou et al. 1996) 
and the submm data for the z $>$ 1 sources are from  
Rowan-Robinson et al. (1993), Isaak et al. (1994), 
Barvainis et al. (1995), Omont et al. (1996b), Eales et al. 1998,
Ivison et al. (1998ab), Lewis et al. (1998), and Kawabe et al. (1998). 
The $3\sigma$ lower limit is shown for the z=1.44
galaxy HR10 (Dey et al. 1998).  
All the radio data are from the Very  Large Array. \\
(b) The curves are the same as in Figure 1a.
The data points are 
the putative identifications of the five submm sources detected
in the Hubble Deep Field by Hughes et al. (1998; 3$\sigma$ lower
limits) and by Richards (1998; filled circles).  Our modeling suggests
that the submm  sources are likely to be at a high redshift ($\rm z
\ge 1.5$). 

\vfill\eject


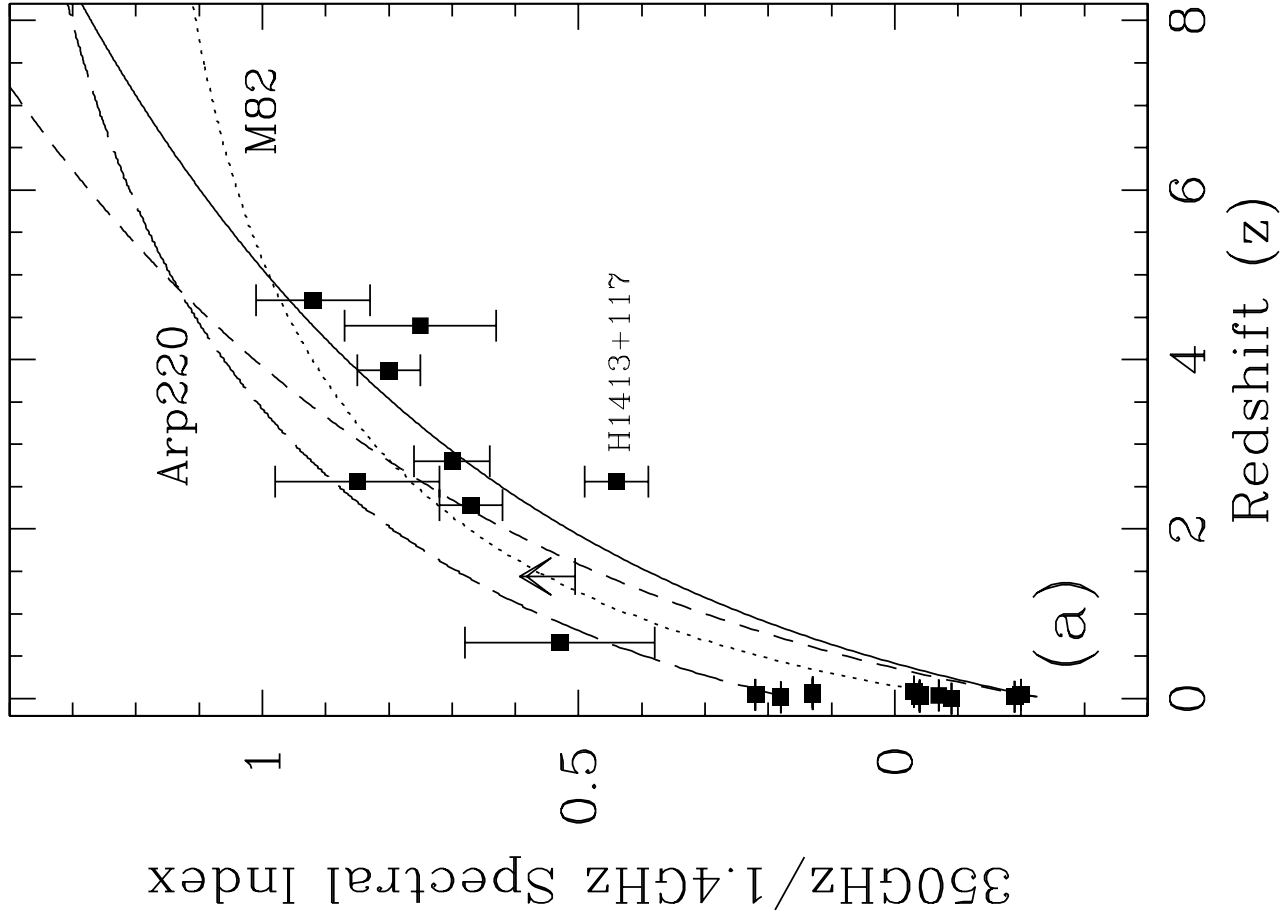
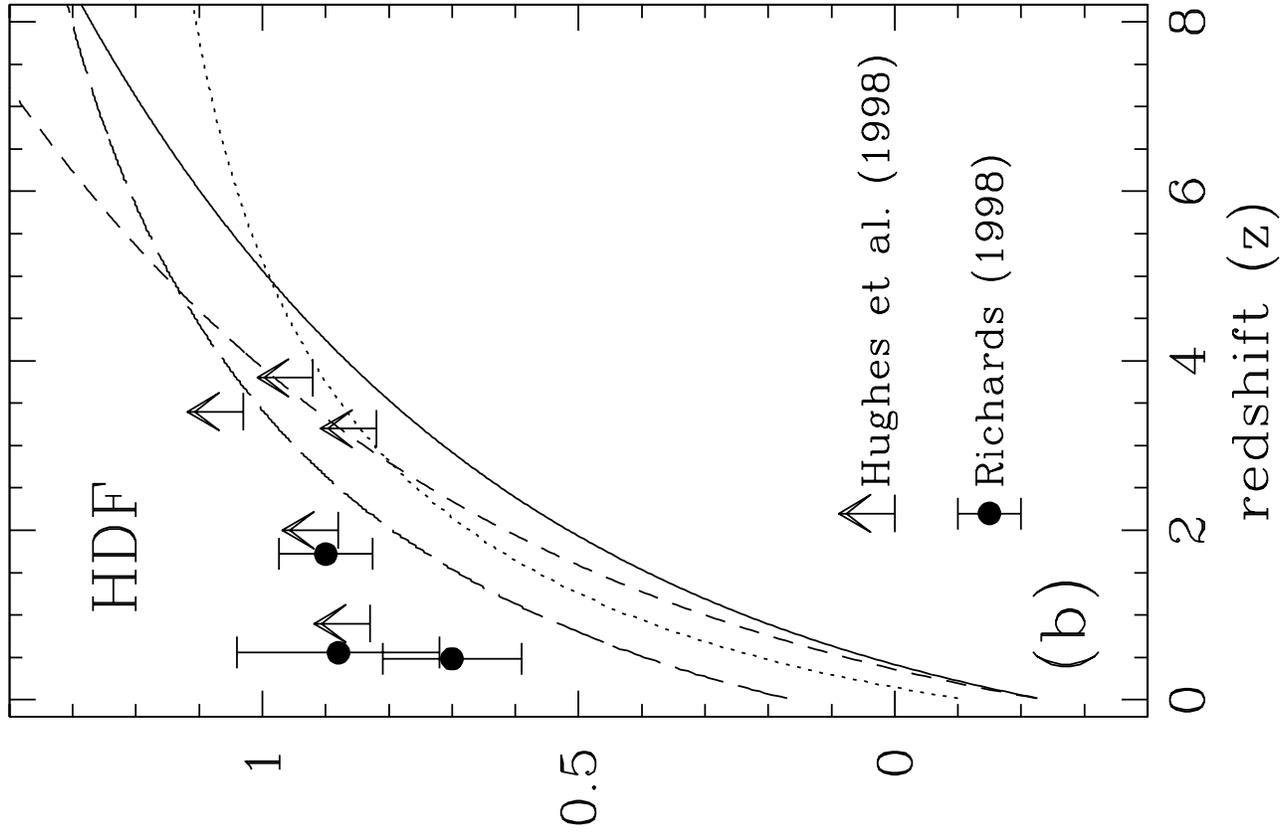
\end{document}